\documentclass[aps,prb,reprint,showpacs,superscriptaddress]{revtex4-1}
\usepackage[dvipdfmx]{graphicx}
\usepackage{bm}
\usepackage{amsmath,amssymb}

\begin{document}

\title{Switching of magnetism via modifying phase shift of quantum-well states by tailoring the interface electronic structure}

\author{Shunsuke Sakuragi}
\email[e-mail: ]{sakuragi@issp.u-tokyo.ac.jp}
\affiliation{ISSP, The University of Tokyo, Kashiwa, Chiba 277-8581, Japan}

\author{Hiroyuki Kageshima}
\affiliation{Department of Physics and Materials Science, Shimane University, Nishikawatsu-cho, Matsue 690-8504, Japan}

\author{Tetsuya Sato}
\affiliation{Department of Applied Physics and Physico-Informatics, Keio University, Hiyoshi, Yokohama 223-0061, Japan}

\date{\today}

\begin{abstract}
We demonstrate control of the magnetism of Pd(100) ultrathin films, which show $d$-electron quantum-well induced ferromagnetism, via modulation of the interface electronic state using density functional calculation. From an analysis based on the phase model, forming the Au/Pd(100) interface induces hybridization of the wave function of $d$-electron quantum-well states, and modulates the term of the scattering phase shift as a function of the reciprocal lattice point. In contrast, forming the Al interface, which has only $s$-electrons at the Fermi energy, cannot modify the scattering phase shift. Our finding indicates the possibility of modifying the phase shift by tailoring the interface electronic states using hybridization of the wave function, and this efficiently changes the density of states near the Fermi energy of Pd films, and the switching between paramagnetism and ferromagnetism occurs based on the condition for ferromagnetism (Stoner criterion). 
\end{abstract}


\maketitle

\section{Introduction}
For the development of novel electronic devices based on the spin degree of freedom of electrons, the switching of the magnetic order of the materials using external fields is an idea having the potential to update the concept of current magnetic memory devices \cite{intro1}. 
Recently, it was experimentally shown that the magnetism and magneto anisotropy of Fe, Co, and Pt ultrathin films\cite{Maruyama,ChibaNatMat,ShimizuPRL} and the proximity-induced magnetism of Pd and Pt ultrathin films, could be modulated using an electric field\cite{ObinataSciRep,YamadaPRL}. 
An experiment involving X-ray magnetic circular dichroism and first-principle calculations indicated that the microscopic origin of the electric field effect on proximity-induced magnetism in Pt ultrathin films, can be explained by a shift of the Fermi level, change in the orbital hybridizations\cite{ObaPRL,YamadaPRL}, and change in the electric quadrupole\cite{MiwaNcom}.

However, in the metal ultrathin film systems used in the electric field effect experiment, the electronic states are modified by the size effect and/or the local lattice distortion of the metals, and the magnetic states are different from those with bulk metals\cite{ShinoharaPRL,GuoScience, ObaPRB, KanaPRB}. 
Especially, the quantum-well states occurring in ultrathin film formed from metal, in particular, modulate the density of states at the Fermi energy $D(\epsilon _F)$ in an oscillatory manner depending on the film thickness. 
The period of oscillation in $D(\epsilon _F)$ is determined by the Fermi wave number of metals\cite{ortegaPRB}. 
The confined electronic states in the quantum-well are described by the phase model\cite{chiang}, 
\begin{equation}\label{phase}
2 k_z d + \Phi = 2 \pi n,
\end{equation}
where $k_z$ is the confined wave number, $d$ is the film thickness, $\Phi$ is the scattering phase shift at the surface and interface, and $n$ is an integer quantum number. 
This equation indicates that the electronic states in the metals are modulated depending on the film thickness. 
In the standpoint of the Stoner criterion, which is the condition required for the appearance of ferromagnetism in transition metals,
\begin{equation}\label{Stoner}
I D(\epsilon_{\rm F}) > 1, 
\end{equation}
where $I$ is the exchange integral, the modifying of $D(\epsilon _F)$ may modulate/induce the ferromagnetic states of 3$d$ and 4$d$ transition metals. 
From this mechanism, the magneto anisotropy of Fe and Co ultrathin films is modulated\cite{DabrowskiPRL,WeberPRL}; in addition, we experimentally/theoretically discovered that ferromagnetism appears in Pd(100) ultrathin films in an oscillatory manner dependent on the film thickness by the quantum-well states\cite{mirbtPRB,niklassonPRB,hongPRB,SakuragiPRB}. 
This quantum-well induced ferromagnetism is modulated by controlling the Fermi level using an electric field\cite{BauerPRB,sunPRB,aiharaJAP} and by modifying the quantum-well states using lattice distortion\cite{SakuragiJMMM,BanAPL,SakuragiPRB2, comment}. 
The mechanism by which these control the magnetism is well understood from the viewpoint of change in the band dispersion.

If the scattering phase shift $\Phi$ is modulated using an external field, it is believed that the magnetism of the entire film is controlled by modulating only the surface and/or interface electronic states of the films. 
However, $\Phi$ is just a parameter in the phase model, and the currently proposed method for controlling $\Phi$ only involves modifying the work function of the materials\cite{MatsudaPRB, ItoPRB}. 
This is because the interpretation of $\Phi$ from the standpoint of electronic states is inarticulate. 
By clarifying the method for controlling $\Phi$, it is possible to develop a novel magnetic switching device by which to control the size effect. 
This will expand the scientific principle upon which is based magnetic switching by band engineering using quantum-well states\cite{TanabeAPL}.


In this paper, we focus on the stacking effect of Au and Al monolayers on the magnetism of Pd(100) ultrathin film. 
Au has $d$-electrons and Al has $s$-electrons around the Fermi energy. 
Based on the above discussion, the stacking of these metals should vary the phase shift $\Phi$, and lead to the modification of the thin-film magnetism. 
The difference between $d$- and $s$-electrons must clarify the role of $\Phi$. 
Actually, we found that these differences lead to a difference of phase shift, and furthermore, that we must regard $\Phi$ as a function of reciprocal space rather than as a simple constant. 
Our findings suggest that the magnetic switching between paramagnetism and ferromagnetism within the metal nano films can be brought on by controlling the dispersion of the band structure originating from quantum-well states, via modulation of the interface electronic states.

\section{Formalism}
The Phase model eq. (\ref{phase}) indicates the conditions under which a standing wave can exist. 
Previous experiments and density functional theoretical (DFT) calculations showed that the period of oscillation by modification of magnetic properties via quantum-well states is expressed as 
\begin{equation}\label{period}
  p = \begin{cases}
    1/(1-k_{Fz}) & (k_{Fz} > 1/2) \\
    1/k_{Fz} & (otherwise)
  \end{cases},
\end{equation}
where $p$ is the period of the oscillation (in thickness of the film $d$) and $k_{Fz}$ is the vertical Fermi wave number, i.e., the Fermi wave number of the confined band \cite{ortegaPRB}. 
Eqs. (\ref{phase}) and (\ref{period}) indicate that the confined wave number $k_z$ reaches Fermi energy periodically depending on the film thickness $d$ \cite{MannaPRB}, and the phase shift $\Phi$ is unrelated to the period of the oscillation. 
The following contents will explain the detail and show the relationship between the $\Phi$ and the shape of the quantum-well band dispersion, which is important to discuss the appearance of the ferromagnetism.

Previous DFT calculations predicted that the $d_{xz, yz}$ electrons are confined at the quantum-well states of Pd(100) films\cite{niklassonPRB, hongPRB, aiharaJAP}. 
These orbital characters express one dimensional dispersion in the in-plane direction. 
For these electrons, the phase shift $\Phi$ of the quantum-well states might depend on $k_x, k_y$, and $\epsilon$. 
Thus, it is necessary to expand the phase model to include $k_x$ and $k_y$ (i.e., the wave number of the in-plane direction $\bm{k_{\parallel}}$) dependency. 
Niklasson {\it et al.} well discussed the relationship between the Phase model and band dispersion of $d_{xz,yz}$ \cite{niklassonPRB}. 
%
%

First, they define $k_z (n) = \left( 2\pi n -\Phi \right)/2d$ from eq. (\ref{period}). 
Then they propose that the $n$-th energy level of the Pd($100$) film quantum well $\epsilon_{\rm QW} \left( n , \bm{k}_{\parallel} \right)$ can be obtained from the Pd bulk energy band structure $\epsilon (k_z, \bm{k}_{\parallel})$ as 
\begin{equation}
\epsilon_{\rm QW} \left( n , \bm{k}_{\parallel} \right)  = \epsilon \left( k_z (n) , \bm{k}_{\parallel} \right), 
\end{equation}
where $\bm{k}_{\parallel} = (k_x, k_y)$ is the in-plane wave vector because the band dispersion of the quantum-well states is a projection of the specific bulk band. 
Then, the binding energy of the quantum-well state $\epsilon_{\rm QW}$ in Pd($100$) can be described by expanding the $\epsilon_{\rm QW}$ around the Fermi energy as 
\begin{equation}\label{expand}
\begin{split}
\epsilon_{\rm QW} \left( n, \bm{k}_{\parallel} \right) 
& \sim \epsilon_{\rm F} + \left[ k_z (n) - k_{{\rm F} z} \right] \frac{\partial \epsilon }{\partial k_z} \left( k_{{\rm F}z}, \bm{k}_{{\rm F}\parallel} \right) 
\\
& + \Delta \bm{k}_{\parallel} \cdot \frac{\partial \epsilon }{\partial \bm{k}_{\parallel}} \left( k_{{\rm F}z}, \bm{k}_{{\rm F} \parallel} \right),
\end{split}
\end{equation}
where $\bm{k}_{{\rm F}{\parallel}}$ the in-plane Fermi wave vector, $\Delta \bm{k}_{\parallel} = \bm{k}_{\parallel} - \bm{k}_{{\rm F}{\parallel}}$, and $\epsilon_{\rm F} = \epsilon \left( k_{{\rm F} z}, \bm{k}_{{\rm F} \parallel} \right)$. 

Considering the Pd bulk band dispersion, the dispersion of $d_{xz,yz}$ has a flat shape around the $\Gamma$ point and zone edges. Therefore, in the Pd($100$) film, the following relation is satisfied: 
\begin{equation}\label{sim}
\frac{\partial \epsilon }{\partial \bm{k}_{\parallel}} \left( k_{{\rm F}z}, \bm{k}_{{\rm F} \parallel} \right) \sim 0.
\end{equation}
According to eq.~(5), this means that $\epsilon_{\rm QW} \left( n, \bm{k}_{\parallel} \right)$ almost coincides with the Fermi energy $\epsilon_{\rm F}$ being independent of $\bm{k}_{\parallel}$, if $k_z(n)$ matches $k_{{\rm F} z}$. 
Then the density of states at the Fermi energy $D(\epsilon_{\rm F})$ diverges and induces ferromagnetism from the standpoint of the Stoner criterion [eq. (\ref{Stoner})]. 
The condition $k_z(n) = k_{{\rm F} z}$ leads to oscillatory appearance of ferromagnetism with the periodicity $p$ depending on the film thickness $d$ as discussed in eq. (\ref{period}). 

Now we extend the above theory. As we will see in the following, the phase shift $\Phi$ generally depends on $\bm{k}_{\parallel}$. If we permit such $\bm{k}_{\parallel}$-dependence for $\Phi$, $k_z(n)$ also depends on $\bm{k}_{\parallel}$. Since $k_z (n,\bm{k}_{\parallel}) = \left( 2\pi n -\Phi (\bm{k}_{\parallel}) \right)/2d$ at present, eq.~(5) must be modified into
\begin{eqnarray}\label{expand2}
\begin{split}
\epsilon_{\rm QW} \left( n, \bm{k}_{\parallel} \right) 
& \sim \epsilon_{\rm F} + \left[ k_z (n,\bm{k}_{{\rm F} \parallel}) - k_{{\rm F} z} \right] \frac{\partial \epsilon }{\partial k_z} \left( k_{{\rm F}z}, \bm{k}_{{\rm F}\parallel} \right) 
\\
& + \Delta \bm{k}_{\parallel} \cdot \frac{\partial \epsilon }{\partial \bm{k}_{\parallel}} \left( k_{{\rm F}z}, \bm{k}_{{\rm F} \parallel} \right) 
\\
&+ \Delta \bm{k}_{\parallel} \cdot \frac{\partial k_z}{\partial \bm{k}_{\parallel}} (n,\bm{k}_{{\rm F} \parallel}) \frac{\partial \epsilon }{\partial k_z} \left( k_{{\rm F}z}, \bm{k}_{{\rm F}\parallel} \right),
\end{split}
\end{eqnarray}
Because of the fourth term on the right hand side, $\epsilon_{\rm QW} \left( n, \bm{k}_{\parallel} \right)$ is no longer the constant for $\bm{k}_{\parallel}$ and the divergence of $D(\epsilon_{\rm F})$ does not often occur, even when $k_z (n,\bm{k}_{{\rm F} \parallel})$ matches $k_{{\rm F} z}$ and eq.~(6) is satisfied. Furthermore, the divergence of $D(\epsilon_{\rm F})$ can be promoted if we appropriately modify the $\bm{k}_{\parallel}$ -dependence of $\Phi$ even when $k_z (n,\bm{k}_{{\rm F} \parallel})$ does not match $k_{{\rm F} z}$ or eq.~(6) is not satisfied. Thus, we can expect that a fine control of the $\bm{k}_{\parallel}$-dependence of phase shift $\Phi$ can reduce the magnetism for magnetic materials as well as induce the magnetism for nonmagnetic materials. 
We note that in eq. (7), the case of $k_{{\rm F} x} = k_{{\rm F} y} =$ 0 is special. From the time reversal symmetry, the first term of Taylor expansion is disappeared when $k_{{\rm F} x} = k_{{\rm F} y} =$ 0. Thus, we have to consider about the higher term of Taylor expansion. 
Nevertheless, we can use same approach [i.e, the considering 4th term of eq. (7)] for the flat shape band structure of QWs. 

To verify this theory, we simulated modification of the interface electronic states of a Pd(100) quantum-well using hybridization of the wave functions by making a transition metal/Pd interface. 
We focused on stacking of the Au monolayer, which is well studied for Pd, and expected interaction of the magnetic $d$-electrons\cite{LiSurfSci,NambaPNAS}. 
In addition, we simulated the Al/Pd(100) system to compare the effects of the $d$-electrons and $s$-electrons using the DFT calculation.

\section{Method}
All DFT calculations were performed with PHASE/0 code\cite{PHASE} using the projector augmented wave\cite{PAW} to the spin-polarized local density approximation reported by Perdew and Wang\cite{LDA}. 
The $60 \times 60 \times 1$ $k$-points and 36 Ry of cutoff energy were used. 
The values of the lattice constant converge to 0.384 nm for fcc bulk Pd (of course the magnetic ground state is nonmagnetic), and we used this value for film-shaped Pd(100). 
Pd(100) ultrathin films express ferromagnetism in an oscillatory manner depending on the layer thickness, and the period of oscillation is 5-6 monolayers of Pd (for example, 4, 9, 10, and 15 monolayers Pd(100) show ferromagnetism: see Ref. \cite{SakuragiPRB2} in Fig. 3). 
In the present study, we calculated the magnetism and electronic states of Au and Al stacked Pd(100) ultrathin films using the slab model in Fig. 1. 
We adjusted the out-of-plane lattice spacing between the stacking layer and Pd to adjust the hybridization of the wave function at the interface of the Pd quantum-well. 
By total energy minimization for the free parameter of the inter-layer distance of Au-Pd and Al-Pd, we found the converged values of 0.205 and 0.163 nm, respectively (see Appendix B). 

Our present obtained lattice constant of 0.384 nm for bulk Pd is smaller than the experimentally observed one of 0.389 nm. 
Basically, the magnetic ground state of Pd is sensitive to lattice constant. 
The Ref. \cite{KanaPRB} shows that the method for volume correction to obtain good agreement with experiment of both lattice constant and magnetic ground state of Pd. 
On the other hand, our previous papers show that our present method using freestanding Pd slab with lattice constant of 0.384 nm obtained from the normal LDA method well reproduces the experimentally observed quantum-well induced ferromagnetism (Ref. \cite{SakuragiPRB,BanAPL,SakuragiPRB2}). 
Thus, in this paper, we use this condition in order to discuss the quantum-well induced ferromagnetism using a simple method.

\begin{figure}
\centering
\includegraphics[width=8.5cm]{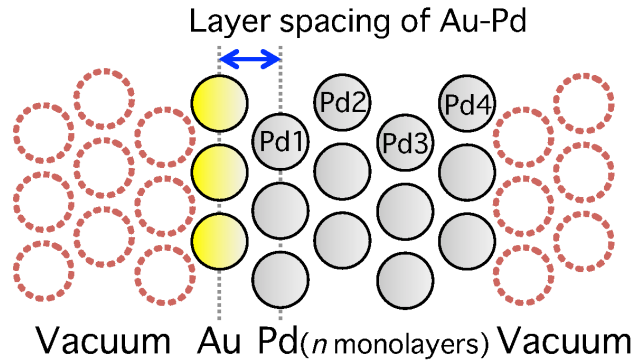}
\caption{\label{slab} 
A schematic image of the slab model of present calculations.  
}
\end{figure}

\section{Results and Discussion}
Figures 2 (a) and (b) show the layer spacing of the Au-Pd dependent magnetic moment of Au (1 monolayer) / Pd (3 monolayers) and Au (1 monolayer) / Pd (4 monolayers), which exhibit non-magnetism and ferromagnetism, respectively, when they are freestanding Pd(100). 
The magnetic moment of the Pd layers is modified by changing the Au-Pd layer spacing. 
Although we also calculated the magnetism of a Au/Pd (2, 5, and 6 monolayers) system, the Pd layers were non-magnetic despite changing of the Au-Pd layer spacing. 
We emphasize that the ferromagnetism was made to appear in Au/Pd (3 monolayers) and disappear in Au/Pd (4 monolayers) by moving the Au layer closer to the Pd film. 
This phenomenon indicates occurrence of increase in the effective film thickness of quantum-well induced ferromagnetism in Pd films by moving of the Au layer closer to Pd, in comparison to the freestanding Pd(100) films. 
In addition, as shown in Fig. 2 (c), not only at the interface of Au/Pd, but also in all layers of the Pd films, the magnetism was modified by modification of the interface electronic states. 
This indicates that change in the interface electronic states affects the entire film.

\begin{figure*}
\includegraphics[width=17.8cm]{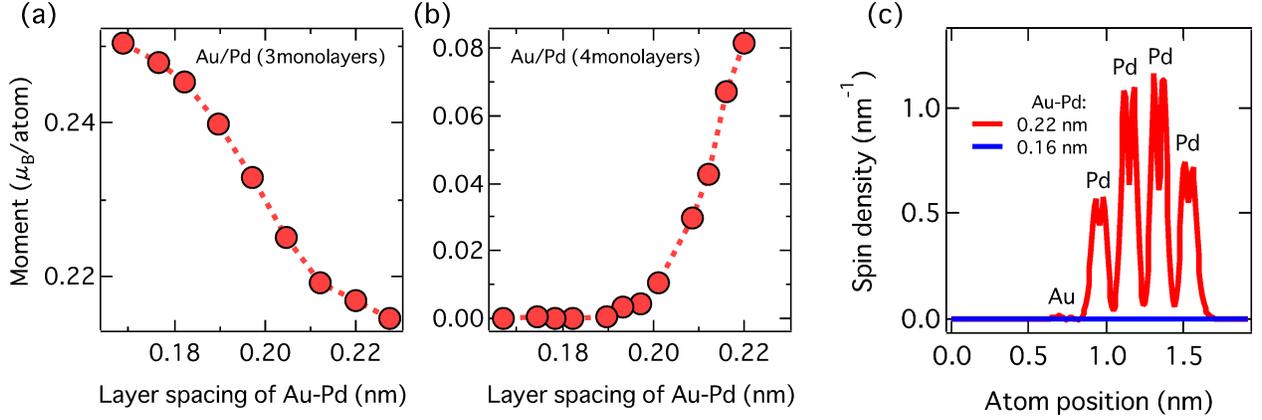}
\caption{\label{mag}
The layer spacing of Au-Pd dependent magnetic moment of (a) Au/Pd(3 monolayer) system and (b) Au/Pd(4 monolayer) system; (c) The layer spacing of Au-Pd dependent spin density in the Au/Pd(4 monolayers) slab. 
}
\end{figure*}

Previous calculation shows that the charge transfer to Pd(100) films can modify the magnetism of Pd\cite{sunPRB,aiharaJAP}. 
By contrast, the present result of the calculation is not explained by a charge transfer from Au to Pd. 
From the previous calculation, a surface carrier density $\Delta \sigma$ greater than $2.8\times 10^{13}$ cm$^{-2}$ is necessary to cancel the ferromagnetic state in four-monolayer Pd(100) film\cite{aiharaJAP}. 
In the present calculation, we estimate the charge transfer between Au and Pd by comparing the charge distribution of pure Au and pure Pd films, and of the layer-spacing-dependent change of the Au/Pd film system. 
Consequently, the occurrence of charge transfer from Au to Pd was less than half the amount necessary to cancel the ferromagnetism of these films.

To investigate the origin of the change in the magnetism of Pd films by the Au stacking effect, we calculated the density of state of Au/Pd film systems depending on the layer spacing of Au-Pd (Figs. 3 a and b). 
The $D(\epsilon _F)$ increases with decrease of the layer spacing of Au-Pd in a Au/Pd-3-monolayer system. 
By contrast, in the Au/Pd-4-monolayer system, decrease in the $D(\epsilon _F)$ by increase in the Au-Pd layer spacing was observed. 
It is known that the magnetic order in metals is determined by the electronic states around the Fermi energy, and that the condition for the appearance of ferromagnetism is described as the Stoner criterion, eq. (\ref{Stoner}). 
From the standpoint of the Stoner criterion, the modification of $D(\epsilon _F)$ causes change in the stability of the ferromagnetic order. 
Thus, the disappearance of ferromagnetism by decrease in the Au-Pd layer spacing in a 4-monolayer Pd system is explained from the standpoint of the Stoner criterion.

\begin{figure}
\centering
\includegraphics[width=9cm]{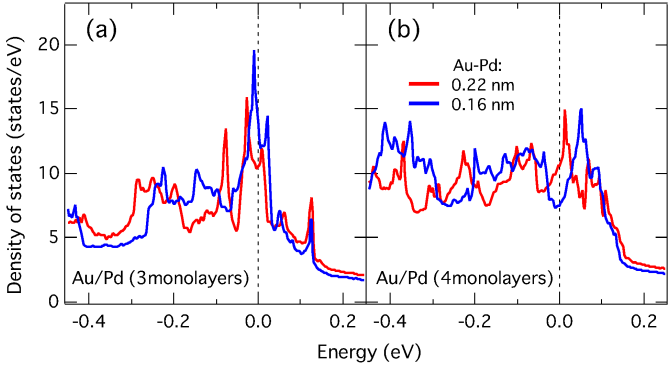}
\caption{\label{dos} 
The layer spacing of Au-Pd dependent the density of states around the Fermi energy of (a) Au/Pd(3 monolayer) system and (b) Au/Pd(4 monolayer) system.
}
\end{figure}

To discuss the change in $D(\epsilon _F)$ from moving the Au layer closer to Pd films, we calculated the band dispersion. 
Figs. 4 (a) and (b) show the band dispersion of the Au/Pd (3 monolayers) with 0.22 and 0.16 nm, respectively, of layer spacing of Au-Pd. 
We verified the effect of the hybridization of the wave functions between Au and Pd by comparing the band dispersion of the freestanding 3-monolayer Pd(100) and Au monolayer (Figs. 4 c and d), respectively. 
The band dispersion that originated from the $d$-electron quantum-well states of Pd are clearly observed on the $\Gamma$-S line defined in Fig. 4 (h) (blue circle in Figs. 4 a to c)\cite{aiharaJAP}. 
In the case of the freestanding Pd film, $\bm{k}_{{\rm F} \parallel} = ( k_{Fx} = 0, k_{Fy} = 0)$ matches with our formalism section II. 
Although the band dispersion that originated from quantum-well states degenerated at the S-point in the freestanding Pd film (green circle in Fig. 4 c), the degeneracy was avoided in the Au/Pd system owing to its broken symmetry (green circle in Figs. 4 a and b).

\begin{figure*}
\includegraphics[width=17.8cm]{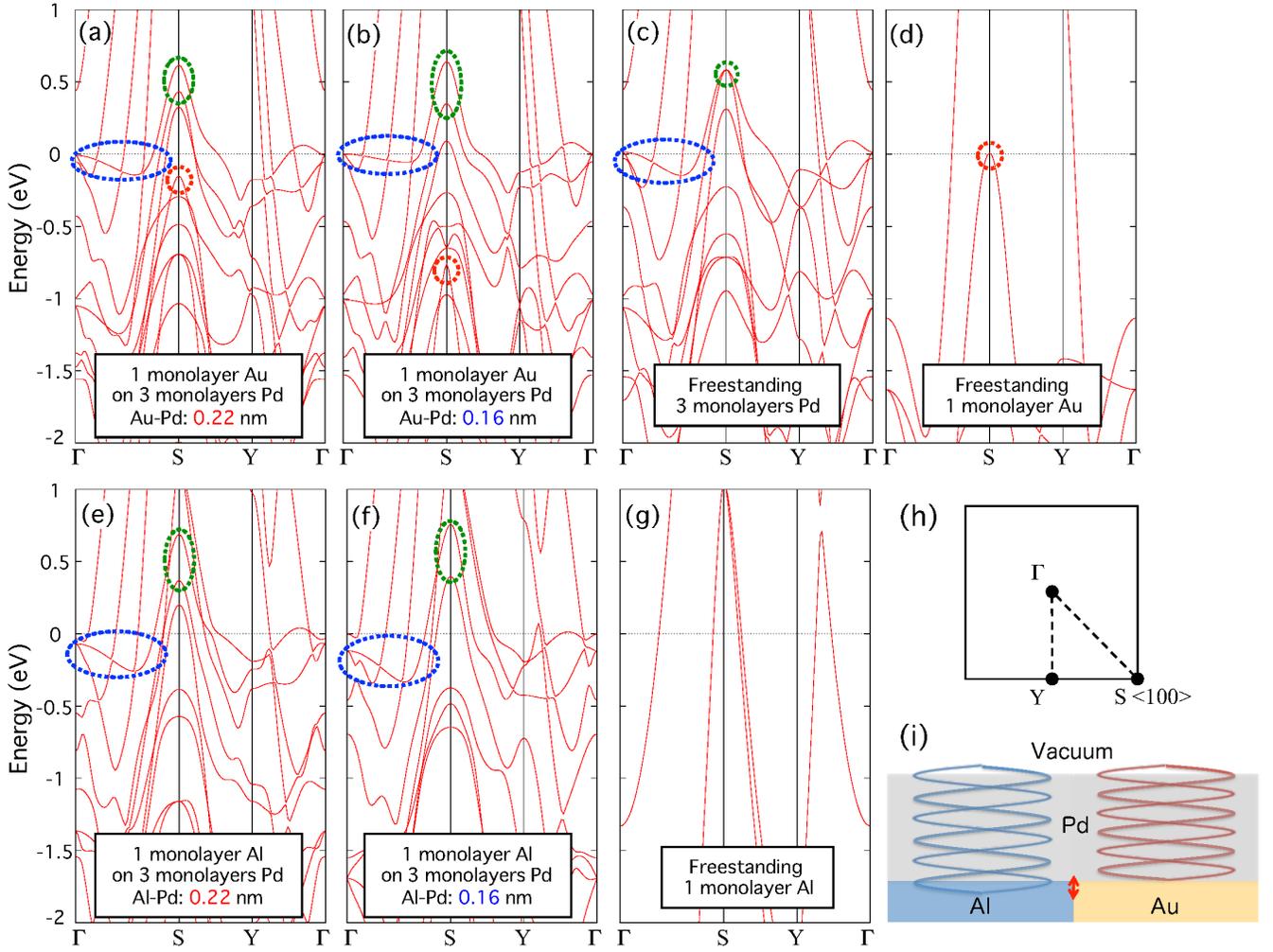}
\caption{\label{band}
The band dispersions of Au/Pd(3 monolayers) system with (a) 0.22 nm and (b) 0.16 nm of the Au-Pd layer spacing; The band dispersion of (c) the freestanding 3 monolayers Pd(100) and (d) 1 monolayer Au; The band dispersions of Al/Pd(3 monolayers) system with (e) 0.22 nm and (f) 0.16 nm of the Al-Pd layer spacing; (g)The band dispersion of the freestanding 1 monolayer Al; (h) Two-dimensional Brillouin zone and the high symmetry points of Pd(100) films; (i) A schematic image of the change in the interface scattering phase shift. }
\end{figure*}

There is a $d$-electron band around the Fermi level at the S-point in monolayer Au (red circle in Fig. 4 d). 
Thus, it is expected that interaction between the Au and $d$-electron quantum-well states of Pd films would occur at the S-point. 
Actually, the value of the band splitting of the Pd quantum-well band at the S-point becomes larger as the Au layer moves closer to Pd (green circle in Figs. 4 a and b). 
In addition, the binding energy of the $d$-electron band of Au (which exists around the S-point in the Au/Pd(100) system; red circle in Figs. 4 a and b; and of which we confirmed the character of the band to see the wave function) becomes larger as Au moves closer to the Pd films. 
These phenomena suggest the occurrence of a hybridization of the wave function between the $d$-electrons of Au and $d$-electron quantum-well states in Pd, and that the hybridization becomes larger as the Au layer moves closer to Pd. 
This hybridization induces modification (flattening) of the band dispersion originating from quantum-well states. 
This phenomenon indicates that there is interaction at the S-point, but no interaction at the $\Gamma$ point, and that lifting of the quantum-well band of Pd film only occurred around the S-point, i.e., the modification of the quantum-well band depends on $\bm{k}_{\parallel}$. According to eq. (\ref{expand2}), the modification of the shape of the quantum-well band dispersion is well explained by the change in the phase shift $\Phi$, and this depends on $\bm{k}_{\parallel}$ indeed. 
This increase of the flatness of the band dispersion increased the $D(\epsilon _F)$ of the system; thus, the change of the density of states in Fig. 3 (a) is well explained.

To evaluate the effect of hybridization of the Au and Pd quantum-well bands, we calculated the wave function. 
Fig. 5 shows the wave function of the quantum-well band of Pd (blue circle in Fig. 4) around the $\Gamma$ and S points drawn on the slab model. 
At the $\Gamma$ point, all the orbital characters are allowed to exist. 
The quantum-well states of Pd originated from confinement of the $d_{xz, yz}$ orbital character\cite{SahaPRB}, and we observed these orbital characters in Pd around the $\Gamma$ point. 
By contrast, around the S-point, the electronic states should be derived mostly from the $d_{x^2 + y^2}$ orbital, from the requirement for symmetry. 
Focusing on the Pd layer adjacent to the Au layer (red dashed square), although the $d_{x^2 + y^2}$ orbital character of Pd was observed around the S-point when the Au layer was far from Pd layer (0.22 nm), $d_{xz, yz}$ orbital-like character was observed when the Au was moved closer to the Pd layer (0.16 nm, blue dashed square). 
In addition, when the distance of Au-Pd layer spacing was 0.16 nm, the wave function of the Au layer (which is hybridized with the quantum-well band of Pd) showed $d_{x^2 + y^2}$-$p_{x, y}$ hybridized orbital character (Fig. 5). 
This phenomenon clearly suggests the occurrence of hybridization between the wave function of Au and the quantum-well band of Pd films around the S-point at the Au/Pd interface. 
This induces modification of the band dispersion of the quantum-well states. 
The evidence of the hybridization between $d$-electrons in Au and Pd quantum-well states is also indicated by the spin density in Fig. 2 (c). 
The Au layer shows a small amount of spin polarization when the quantum-well induced ferromagnetism appears in Pd(100). 
Au is a typical nonmagnetic transition metal; thus, the spin polarization in the Au layer of the Au/Pd(100) system is caused by the hybridization between the Au and magnetic Pd $d$-electrons.

\begin{figure*}
\includegraphics[width=17.8cm]{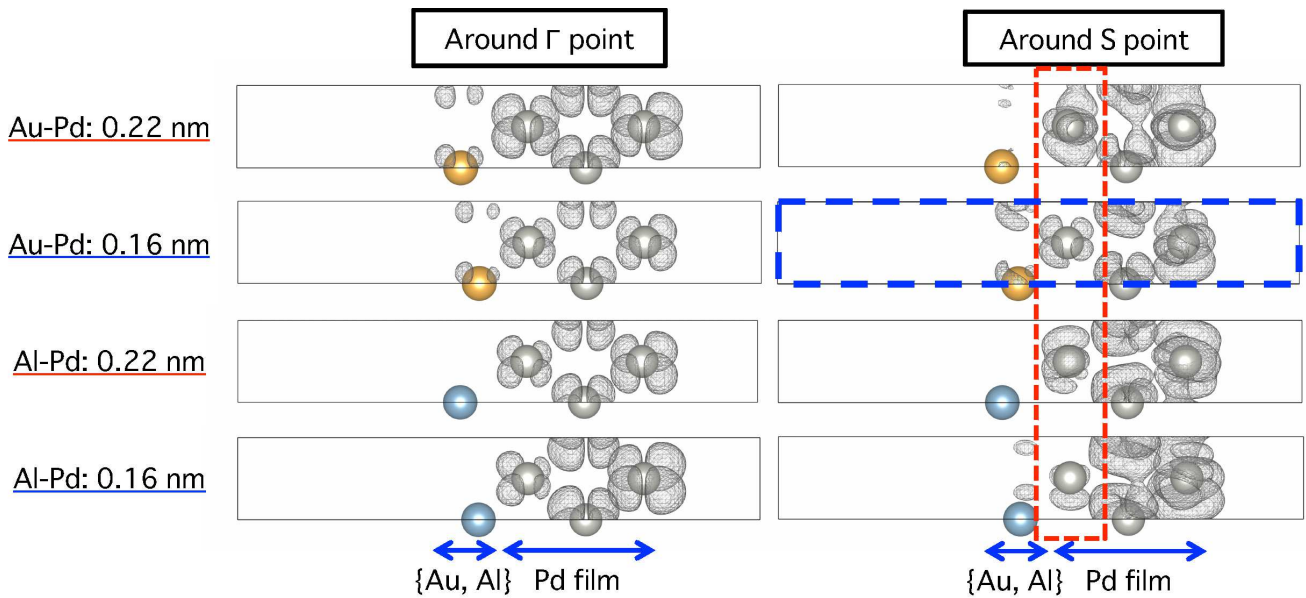}
\caption{\label{charge}
Layer spacing of (Au, Al) - Pd films dependent wave function of the quantum-well states of Pd around $\Gamma$ and S-point. 
}
\end{figure*}

This interaction of the quantum-well states depending on the position of the reciprocal lattice space is caused by a stacking layer having $d$-electrons (i.e., Au). 
In fact, the same effect is observed in a system in which an Fe monolayer is stacked on 4-monolayer Pd(100). 
The layer spacing of Fe-Pd converges to 0.169 nm, and all of the Pd layers show spin polarization at that time. 
In contrast, only the Pd layer adjacent to the Fe layer shows ferromagnetism (proximity-induced magnetism) when the layer spacing of Fe-Pd is 0.192 nm (the lattice constant of Pd). 
Thus, our finding is that magnetic change generally occurs when a $d$-electron transition metal is stacked on the Pd(100) films. 
Here, we also stacked Al (which has only $s$-electrons at the Fermi energy), on the 3-monolayer Pd(100) to compare it with $d$-transition metals. 
Although the Fermi level of Pd is modified by the charge transfer from Al which erase the ferromagnetism, the shape of the band dispersion is not modified by changing the Al-Pd layer spacing (Figs. 4 e to g). 
The symmetry of the orbital character of the quantum-well states of Pd films are also not modified in relation to the Al-Pd layer spacing around the S-point (Fig. 5). 
This indicates that there is no hybridization between the Al $s$-band and $d$-electron quantum-well band of the Pd films. 
The final term of eq. (\ref{expand2}) is zero in the case of Al stacking system, on the other hand, that of Au stacking system is non-zero. 
Therefore, it is suggested that adsorption of a transition metal having $d$-electrons is essentially important to control quantum-well induced ferromagnetism using modulation of the shape of the band dispersion of the quantum-well bands, i.e., the effect of the modification of $\Phi$ depending on $\bm{k}_{\parallel}$. 
From eq. (\ref{phase}), this modification is interpreted that the change in the effective thickness of the quantum-well occurs only around the Au-stacked S-point (Fig. 4 i).

According to the phase model in eq. (\ref{phase}) and our present obtained eq. (\ref{expand2}), enhancement of the flatness of the band dispersion originating from the quantum-well states around the zone edge, indicates the occurrence of modification in the phase shift $\Phi$, which depends on $\bm{k}_{\parallel}$ indeed. 
Our present DFT calculations show that the $d$-electron quantum-well-induced ferromagnetism is controlled by change at the interface electronic states via stacking of an overlayer having $d$-electrons. 
Hitherto, the term $\Phi$ was just a parameter for describing the quantum-well states in the real space. 
Our present DFT calculation suggests a mechanism by which to control the term of $\Phi$ as a function of the position of the reciprocal lattice point. 
Thus, our finding extends the interface effect of the quantum-well states, which was only understood in real space, from the standpoint of the reciprocal lattice space. 
This mechanism indicates the possibility of controlling the electronic states in whole films using the effect from hybridization of wave functions because of modification of the interface electronic states of the metal-nano film structures.

We note that our previous studies show that the model of freestanding Pd slab, which is used also in this study, well explain the quantum-well induced ferromagnetism in Pd films \cite{SakuragiPRB, BanAPL, SakuragiPRB2}. In addition, we also see the stacking effect for the quantum-well states discussed in this paper in the satisfyingly thick film to perform the experiment. 
Thus, we would expect the control of magnetism predicted here to be experimentally observable in a Au/Pd system on a piezo substrate. 
In this system, the change in the out-of-plane lattice constant occurs via occurrence of the piezo effect on the in-plane lattice constant, by which modification of the Au-Pd layer spacing is realized. 
The previous experiment and DFT calculation showed evidence of control of the magnetic moment in pure Pd(100) films via lattice distortion from a ferroelectric BaTiO$_3$ substrate \cite{BanAPL}. 
The lattice expansion induced by the lattice distortion of the substrate causes narrowing of the density of states of the bulk band, and changes its magnetic properties\cite{MoruzziPRB, PhysProc, SakuragiPRB2, ShigaJPSJ, EkmanPRB, MankovskyPRB}. 
However, in the pure Pd(100) ultrathin films on a BaTiO$_3$ substrate, the amount of change in the magnetic moment was only 5 \%. 
Our present DFT calculation suggests that modification of the band dispersion that originates from the quantum-well states will produce a synergistic effect able to alter the magnetism by lattice distortion using the Au/Pd/piezo-substrate heterostructure. 
Using this mechanism, it is expected that nonmagnetic to ferromagnetic switching could be provided by applying lattice distortion in the quantum-well-induced ferromagnetism.

In the case of the quantum confinement of the Pd $d_{xz, yz}$ electrons, which we discussed in this paper, the hybridization between the $d$-electron wave function of the stacking layer and Pd at the zone edge (S-point) is intrinsically important for producing a flat band. 
Because typical 3, 4, 5$d$ transition metals forming fcc structure have $d$-electron bands around the S-point, this effect generally occurs by stacking of these transition metals on Pd(100) films. 
In addition, if the $d$-electron band of a stacking layer appears near the Fermi energy at the S-point, the magnetic change might be clearly observed, as indicated in eq. (\ref{expand2}).

\section{Conclusions}
In conclusion, we investigated the magnetic change in Pd(100) ultrathin films with quantum-well induced ferromagnetism caused by the stacking effect of the transition metal using the DFT calculation. 
When a Au layer with $d$-electrons was stacked on the $d$-electron quantum-well, modification of the quantum-well induced ferromagnetism was observed. 
This phenomenon is explained by modification of the term of the interface scattering phase shift $\Phi$ as a function of the reciprocal lattice space via hybridizing of the wave functions between the band dispersion of the stacking $d$-electron and $d$-electron quantum-well states. 
In the case of confined $d_{xz, yz}$ electrons, modification of $\Phi$ only occurred around the zone edge, and the dispersion shape flattened (i.e., the density of the states was modified). 
Contrastingly, we observed that the stacking of Al, which contains only $s$-electrons around Fermi energy, cannot modify the shape of the band dispersion of the $d$-electron quantum-well states because there is no hybridization between the $s$- and $d$-electrons. 
Our findings suggest a mechanism for controlling magnetism using modification of the interface electronic states in metal-nano structures. 
This mechanism could be extended to other magnetic materials, creating the possibility of tailoring magnetic materials by appropriate electronic engineering of the interface structure.

\section*{ACKNOWLEDGMENTS}
We would like to thank to E. Minamitani, M. Matsubara, S. Shin, T. Kondo, H. Wadati, K. Okazaki, and K. Kuroda for valuable discussions. 
The computation in this work has been done using the facilities of the Supercomputer Center, the Institute for Solid State Physics, the University of Tokyo. 
This work was supported by Japan Society for the Promotion of Science Grants-in-Aid for Scientific Research (KAKENHI) Grant Number 15H01998 and 19K05199. 

\bibliography{ref}

\appendix
\section{Au-Pd distance dependence of quantum-well states}
In Fig. 4, we show the change in the quantum-well states of Pd(100) ultrathin film with 0.22 nm and 0.16 nm of Au-Pd layer spacing. In Fig. 6, we show the band dispersion of Au/Pd(3 monolyers) system with various Au-Pd layer spacing. We can see the gradually changes of the quantum-well band dispersion of Pd films depending on the Au-Pd layer spacing. 
In addition, we note that we can see the same result in Au/Pd(4 monolayers) system. 

\begin{figure*}
\includegraphics[width=17.8cm]{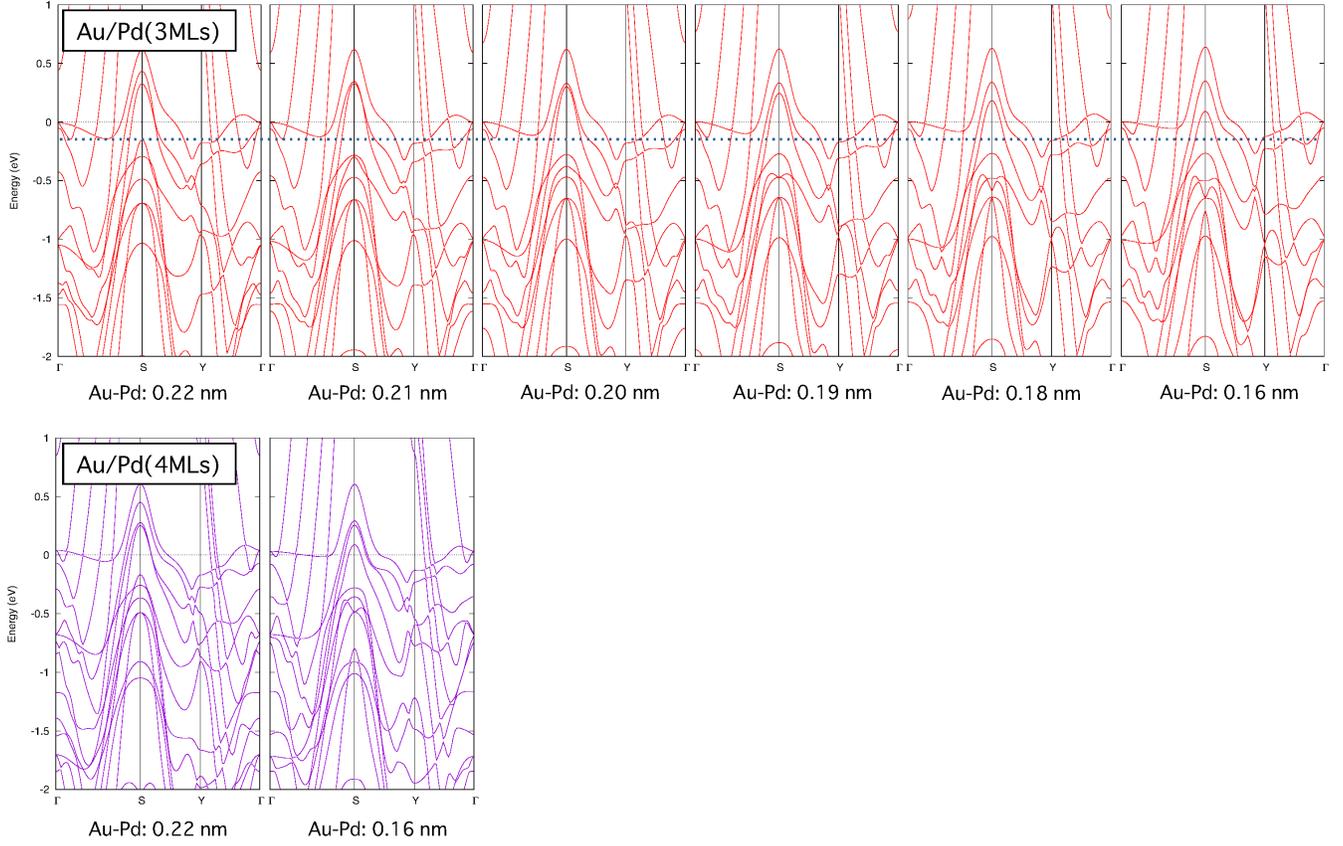}
\caption{\label{band_d}
The band dispersions of Au/Pd(3 and 4 monolayers: MLs) system with from 0.22 nm to 0.16 nm of the Au-Pd layer spacing. 
The lattice constant of Pd is fixed to the bulk converged value. 
}
\end{figure*}

\section{Full structural relaxations of Au/Pd(2-6 monolayers) and Al/Pd(2-6 monolayers)}
To discuss the magnetic ground states, we should consider the film structure because the magnetism in Pd is sensitive to the lattice parameter. 
Thus, we perform the full structural relaxations of Au/Pd(2-6 monolayers) and Al/Pd(2-6 monolayers) systems (Table 1 and 2). As a result, the in-plane lattice constants are not changed, but the out-of-plane lattice constants are changed. 
We note that there is no Pd thickness dependence of layer spacing of Au-Pd and Al-Pd. 
After the full structural relaxations, the stability of magnetic states are little bit modified, and we cannot see the magnetic moment in Au/Pd(4 monolayers) system. 
On the other hand, we see the spontaneous magnetization of Pd in Au/Pd(3 monolayers) system even in the lattice parameter obtained from the full structural relaxations. 
Then, we fixed Pd-Pd layer spacing to the values obtained from the full structural relaxations, and we checked the Au-Pd layer spacing dependence of magnetism for Au/Pd(3 monolayer) system (Fig. 7). 
As a result, we see the same behaviour as Fig. 2(a). 
We note that we cannot see the ferromagnetic order in Al/Pd(3 monolayers) system when we modified the Al-Pd distance. Therefore, our present summary is satisfied even in the system with full structural relaxations.

\begingroup
\squeezetable
\begin{table*}
  \begin{tabular}{|c||c|c|c|c|c|c|c|} \hline
 & Moment ($\mu _B$/Pd) 				& Au-Pd1(nm)	& Pd1-Pd2(nm)	& Pd2-Pd3(nm)	& Pd3-Pd4(nm)	& Pd4-Pd5(nm)	& Pd5-Pd6(nm) \\ \hline\hline
    Au/Pd(2 MLs) & 0.00					& 0.205		&0.189		& 			&			&			&			\\ \hline
    Au/Pd(3 MLs) & 0.23					& 0.205		&0.194		&0.189		&			&			&			\\ \hline
    Au/Pd(4 MLs) & 0.00					& 0.204		&0.193		&0.191		&0.189		&			&			\\ \hline
    Au/Pd(5 MLs) & 0.00					& 0.205		&0.194		&0.192		&0.191		&0.189		&			\\ \hline
    Au/Pd(6 MLs) & 0.00					& 0.205		&0.194		&0.194		&0.192		&0.192		&0.189		\\ \hline
  \end{tabular}
  \caption{
  Out-of-plane lattice constant of Au/Pd(2-6 monolayers: MLs). There is no change in the in-plane lattice constants from bulk converged value. 
  }
  \label{tb:slash}
\end{table*}
\endgroup
\begingroup
\squeezetable
\begin{table*}
  \begin{tabular}{|c||c|c|c|c|c|c|c|} \hline
 & Moment ($\mu _B$/Pd) 				& Al-Pd1(nm)	& Pd1-Pd2(nm)	& Pd2-Pd3(nm)	& Pd3-Pd4(nm)	& Pd4-Pd5(nm)	& Pd5-Pd6(nm) \\ \hline\hline
    Al/Pd(2 MLs) &  0.00					& 0.162		&0.197		& 			&			&			&			\\ \hline
    Al/Pd(3 MLs) &  0.00					& 0.162		&0.202		& 0.190		&			&			&			\\ \hline
    Al/Pd(4 MLs) &  0.00					& 0.162		&0.201		&0.193		&0.189		&			&			\\ \hline
    Al/Pd(5 MLs) &  0.00					& 0.162		&0.200		&0.194		&0.192		&0.188		&			\\ \hline
    Al/Pd(6 MLs) &  0.00					& 0.162  		&0.201 		&0.193 		&0.192 		&0.191 		&0.188 		\\ \hline
  \end{tabular}
  \caption{
    Out-of-plane lattice constant of Al/Pd(2-6 monolayers: MLs). There is no change in the in-plane lattice constants from bulk converged value.  }
  \label{tb:slash2}
\end{table*}
\endgroup

\begin{figure}
\centering
\includegraphics[width=9cm]{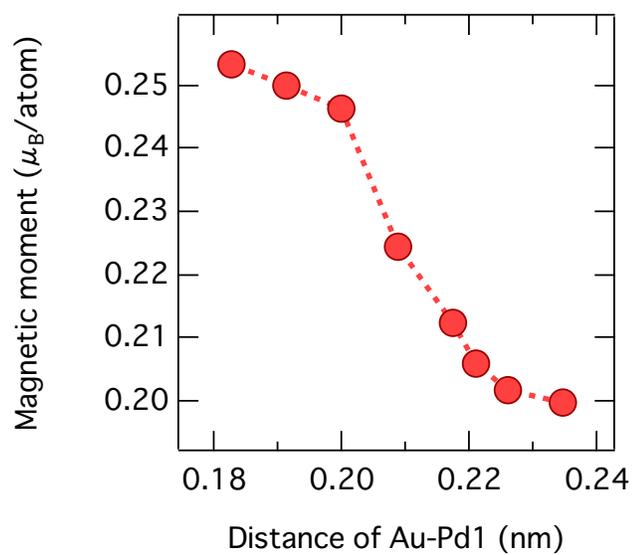}
\caption{\label{relaxed} 
The layer spacing of Au-Pd dependent the magnetic moment of Au/Pd(3 monolayer) system with Pd structure obtained from full structural relaxations.
}
\end{figure}

\end{document}